\documentclass[aps,prd,twocolumn,superscriptaddress]{revtex4-1}

\usepackage{graphicx}
\usepackage[caption=false]{subfig}

\begin{document}



\title{Search for Astrophysical Tau Neutrinos in Three Years of
  IceCube Data}




\affiliation{III. Physikalisches Institut, RWTH Aachen University, D-52056 Aachen, Germany}
\affiliation{Department of Physics, University of Adelaide, Adelaide, 5005, Australia}
\affiliation{Dept.~of Physics and Astronomy, University of Alaska Anchorage, 3211 Providence Dr., Anchorage, AK 99508, USA}
\affiliation{CTSPS, Clark-Atlanta University, Atlanta, GA 30314, USA}
\affiliation{School of Physics and Center for Relativistic Astrophysics, Georgia Institute of Technology, Atlanta, GA 30332, USA}
\affiliation{Dept.~of Physics, Southern University, Baton Rouge, LA 70813, USA}
\affiliation{Dept.~of Physics, University of California, Berkeley, CA 94720, USA}
\affiliation{Lawrence Berkeley National Laboratory, Berkeley, CA 94720, USA}
\affiliation{Institut f\"ur Physik, Humboldt-Universit\"at zu Berlin, D-12489 Berlin, Germany}
\affiliation{Fakult\"at f\"ur Physik \& Astronomie, Ruhr-Universit\"at Bochum, D-44780 Bochum, Germany}
\affiliation{Physikalisches Institut, Universit\"at Bonn, Nussallee 12, D-53115 Bonn, Germany}
\affiliation{Universit\'e Libre de Bruxelles, Science Faculty CP230, B-1050 Brussels, Belgium}
\affiliation{Vrije Universiteit Brussel, Dienst ELEM, B-1050 Brussels, Belgium}
\affiliation{Dept.~of Physics, Chiba University, Chiba 263-8522, Japan}
\affiliation{Dept.~of Physics and Astronomy, University of Canterbury, Private Bag 4800, Christchurch, New Zealand}
\affiliation{Dept.~of Physics, University of Maryland, College Park, MD 20742, USA}
\affiliation{Dept.~of Physics and Center for Cosmology and Astro-Particle Physics, Ohio State University, Columbus, OH 43210, USA}
\affiliation{Dept.~of Astronomy, Ohio State University, Columbus, OH 43210, USA}
\affiliation{Niels Bohr Institute, University of Copenhagen, DK-2100 Copenhagen, Denmark}
\affiliation{Dept.~of Physics, TU Dortmund University, D-44221 Dortmund, Germany}
\affiliation{Dept.~of Physics and Astronomy, Michigan State University, East Lansing, MI 48824, USA}
\affiliation{Dept.~of Physics, University of Alberta, Edmonton, Alberta, Canada T6G 2E1}
\affiliation{Erlangen Centre for Astroparticle Physics, Friedrich-Alexander-Universit\"at Erlangen-N\"urnberg, D-91058 Erlangen, Germany}
\affiliation{D\'epartement de physique nucl\'eaire et corpusculaire, Universit\'e de Gen\`eve, CH-1211 Gen\`eve, Switzerland}
\affiliation{Dept.~of Physics and Astronomy, University of Gent, B-9000 Gent, Belgium}
\affiliation{Dept.~of Physics and Astronomy, University of California, Irvine, CA 92697, USA}
\affiliation{Dept.~of Physics and Astronomy, University of Kansas, Lawrence, KS 66045, USA}
\affiliation{Dept.~of Astronomy, University of Wisconsin, Madison, WI 53706, USA}
\affiliation{Dept.~of Physics and Wisconsin IceCube Particle Astrophysics Center, University of Wisconsin, Madison, WI 53706, USA}
\affiliation{Institute of Physics, University of Mainz, Staudinger Weg 7, D-55099 Mainz, Germany}
\affiliation{Universit\'e de Mons, 7000 Mons, Belgium}
\affiliation{Technische Universit\"at M\"unchen, D-85748 Garching, Germany}
\affiliation{Bartol Research Institute and Dept.~of Physics and Astronomy, University of Delaware, Newark, DE 19716, USA}
\affiliation{Dept.~of Physics, Yale University, New Haven, CT 06520, USA}
\affiliation{Dept.~of Physics, University of Oxford, 1 Keble Road, Oxford OX1 3NP, UK}
\affiliation{Dept.~of Physics, Drexel University, 3141 Chestnut Street, Philadelphia, PA 19104, USA}
\affiliation{Physics Department, South Dakota School of Mines and Technology, Rapid City, SD 57701, USA}
\affiliation{Dept.~of Physics, University of Wisconsin, River Falls, WI 54022, USA}
\affiliation{Oskar Klein Centre and Dept.~of Physics, Stockholm University, SE-10691 Stockholm, Sweden}
\affiliation{Dept.~of Physics and Astronomy, Stony Brook University, Stony Brook, NY 11794-3800, USA}
\affiliation{Dept.~of Physics, Sungkyunkwan University, Suwon 440-746, Korea}
\affiliation{Dept.~of Physics, University of Toronto, Toronto, Ontario, Canada, M5S 1A7}
\affiliation{Dept.~of Physics and Astronomy, University of Alabama, Tuscaloosa, AL 35487, USA}
\affiliation{Dept.~of Astronomy and Astrophysics, Pennsylvania State University, University Park, PA 16802, USA}
\affiliation{Dept.~of Physics, Pennsylvania State University, University Park, PA 16802, USA}
\affiliation{Dept.~of Physics and Astronomy, Uppsala University, Box 516, S-75120 Uppsala, Sweden}
\affiliation{Dept.~of Physics, University of Wuppertal, D-42119 Wuppertal, Germany}
\affiliation{DESY, D-15735 Zeuthen, Germany}

\author{M.~G.~Aartsen}
\affiliation{Department of Physics, University of Adelaide, Adelaide, 5005, Australia}
\author{K.~Abraham}
\affiliation{Technische Universit\"at M\"unchen, D-85748 Garching, Germany}
\author{M.~Ackermann}
\affiliation{DESY, D-15735 Zeuthen, Germany}
\author{J.~Adams}
\affiliation{Dept.~of Physics and Astronomy, University of Canterbury, Private Bag 4800, Christchurch, New Zealand}
\author{J.~A.~Aguilar}
\affiliation{Universit\'e Libre de Bruxelles, Science Faculty CP230, B-1050 Brussels, Belgium}
\author{M.~Ahlers}
\affiliation{Dept.~of Physics and Wisconsin IceCube Particle Astrophysics Center, University of Wisconsin, Madison, WI 53706, USA}
\author{M.~Ahrens}
\affiliation{Oskar Klein Centre and Dept.~of Physics, Stockholm University, SE-10691 Stockholm, Sweden}
\author{D.~Altmann}
\affiliation{Erlangen Centre for Astroparticle Physics, Friedrich-Alexander-Universit\"at Erlangen-N\"urnberg, D-91058 Erlangen, Germany}
\author{T.~Anderson}
\affiliation{Dept.~of Physics, Pennsylvania State University, University Park, PA 16802, USA}
\author{I.~Ansseau}
\affiliation{Universit\'e Libre de Bruxelles, Science Faculty CP230, B-1050 Brussels, Belgium}
% ...

\date{\today}

\collaboration{IceCube Collaboration}
\noaffiliation


\begin{abstract}
The IceCube Neutrino Observatory has observed a diffuse flux of
TeV-PeV astrophysical neutrinos at 5.7$\sigma$ significance from an all-flavor
search. The direct detection of tau neutrinos in this flux has yet to occur. Tau neutrinos become distinguishable from other flavors
in IceCube at energies above a few hundred TeV, when the cascade from
the tau neutrino charged current interaction becomes resolvable from
the cascade from the tau lepton decay. This paper presents results
from the first dedicated search for tau neutrinos with energies between 214
TeV and 72 PeV in the full IceCube detector. The analysis searches for IceCube optical sensors that observe two
separate pulses in a single event - one from the tau neutrino interaction, and a second from the tau decay.
No candidate events were observed in three years of IceCube data. For the first time, a
differential upper limit on astrophysical tau neutrinos is derived around
the PeV energy region, which is nearly three orders of magnitude lower in
energy than previous limits from dedicated tau neutrino searches.
\end{abstract}


\maketitle

\section{Introduction}
The IceCube Neutrino observatory has announced a significant detection
of a diffuse astrophysical neutrino flux above 30
TeV~\cite{icecube2013evidence, aartsen2014observation}. 
The source of this flux is as yet unknown, with
possible candidates including cosmic ray acceleration in active
galactic nuclei, gamma ray bursts, and supernova remnants. Assuming standard 3-flavor
oscillations and a most commonly considered $\nu_e$ : $\nu_{\mu}$ : $
\nu_{\tau}$ = 1 : 2 : 0 flux from pion decay at the source, 
the neutrinos detected in IceCube should be divided almost equally into all
three flavors~\cite{learned1995detecting}. Other flavor compositions
at the source ranging from 1 : 0 : 0 to 0 : 1 : 0 are possible for
dominant processes such as neutron decay~\cite{Anchordoqui200442},
energy loss of pions and muons before decay in environment with strong magnetic fields or high matter
density~\cite{PhysRevD.58.123005, PhysRevLett.95.181101,
  PhysRevD.74.063009, PhysRevD.75.123005} and muon acceleration~\cite{klein2013muon}. Though those scenarios
result in non-universal flavor ratios at Earth, they all predict 
significant fluxes of tau neutrinos after averaged oscillations by propagation over
astronomical distances~\cite{anchordoqui2014cosmic,
  PhysRevLett.114.171102, Aartsen:2015knd}. Above PeV energies, the Earth becomes opaque to
electron and muon neutrinos, while the tau neutrino flux is
regenerated through subsequent tau lepton decays to neutrinos~\cite{halzen1998tau}.  
Tau neutrino background from the atmosphere is expected to be negligible at high energies, 
with only a small contribution from the decay of charmed
mesons~\cite{PhysRevD.78.043005}. Therefore the detection of tau neutrinos at
high energies would both give new information about the astrophysical
flux as well as serving as an additional confirmation of the astrophysical origin of the high energy diffuse
neutrino signal. 
Two recent flavor ratio analyses of IceCube high energy
neutrino events were consistent with equal fractions of all
flavors in IceCube, though with large
uncertainty~\cite{PhysRevLett.114.171102, Aartsen:2015knd}. 
Neither flavor ratio analysis included a dedicated tau neutrino
identification algorithm, which would improve the measurement of
astrophysical neutrino flavor ratios. Precise measurement of
astrophysical neutrino flavor content at Earth 
will shed light on the emission mechanisms at the source, test the
fundamental properties of neutrinos over extremely long baselines
and better constrain new physics models which predict significant deviations from
equal fractions of all flavors~\cite{PhysRevLett.90.181301,baerwald2012neutrino,
  PhysRevD.62.103007, PhysRevLett.92.011101, PhysRevD.81.013006,
  PhysRevD.72.065009, PhysRevD.72.065019,  PhysRevLett.115.161303,
  PhysRevLett.115.161302, Palladino:2015vna}.

\begin{figure}
\includegraphics[width=0.45\textwidth]{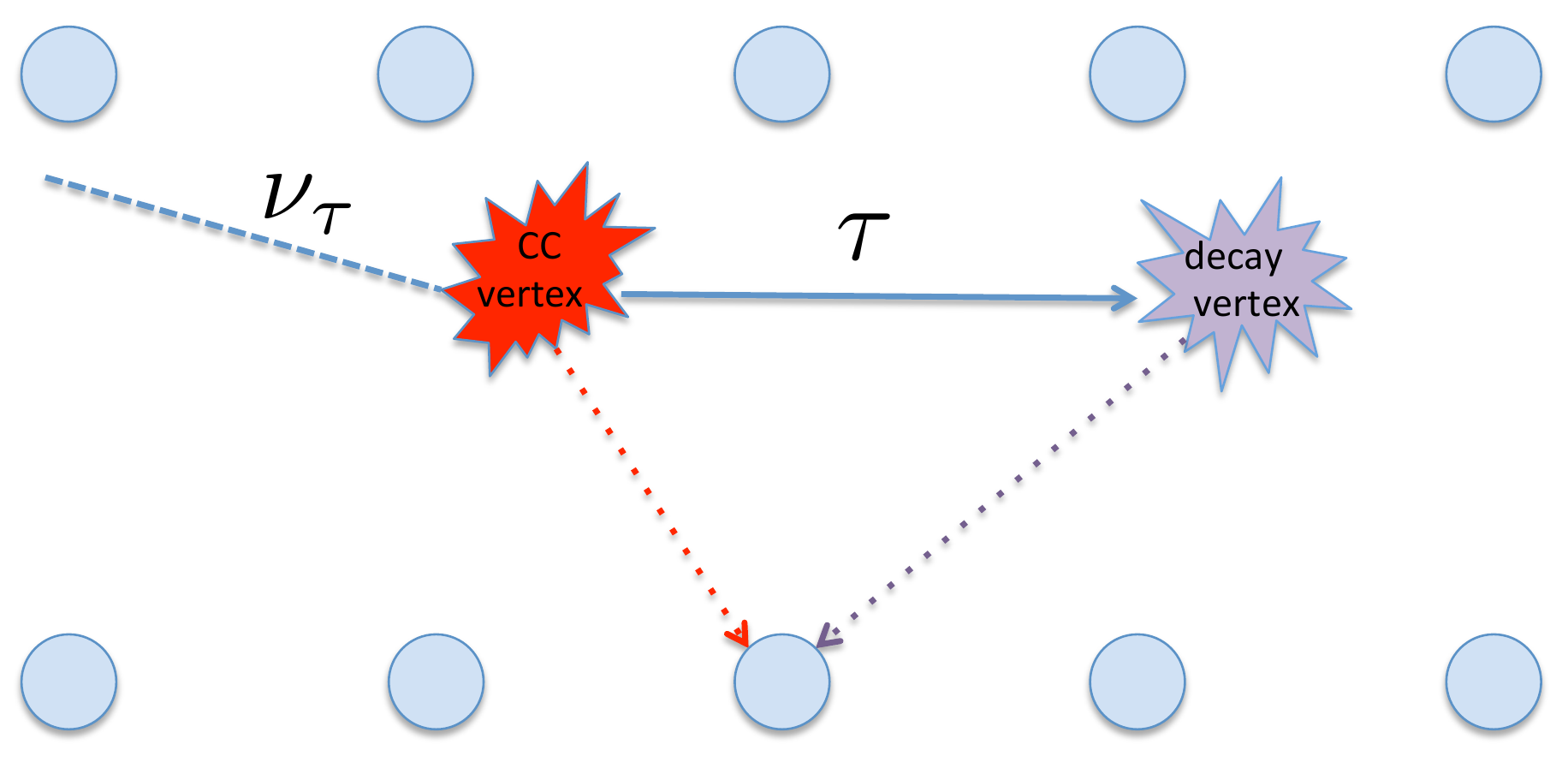}
 \caption{Sketch of a $\nu_{\tau}$ undergoing a charged current
   interaction and producing a double bang signature in IceCube. The
   blue circles represent IceCube photon sensors buried in the
   ice. This figure is not drawn to scale.}
  \label{TauDP}
\end{figure}

Most neutrino interactions in IceCube have one of two event
topologies: tracks from charged current (CC) interactions of muon
neutrinos, and cascades (or showers) from CC interactions of electron
and low energy tau neutrinos and neutral current (NC) interactions of
all flavors. As the average tau decay length roughly scales as
5~cm/TeV, at energies above a few hundred TeV, the tau lepton produced in a tau
neutrino CC interaction would have a decay length sufficiently long
that the CC interaction of the tau neutrino and the subsequent decay
of the tau lepton could be resolved by IceCube sensors. 
There is an 83\% chance that the tau lepton decays to electrons or
hadrons, producing a second cascade. 
This double cascade signal is called a ``double bang''~\cite{learned1995detecting}; an event sketch is shown in Figure~\ref{TauDP}. 

A previous IceCube search for high energy tau neutrinos was published
using data from the partially completed
detector~\cite{abbasi2012search}. 
Since a fully contained double bang is unlikely to be contained
in the small size of the incomplete detector, 
the search was optimized for partially contained double bangs. The search led to a null result and was in fact more sensitive to
electron and muon flavor neutrinos than to tau neutrinos. 

There is a 17\% chance of the tau decaying to a muon, producing an outgoing
track. Therefore, another possible search method is to look for a
track that abruptly brightens along its length as the muon produces
light more efficiently than the parent tau~\cite{cowen2007tau}. 
Tau neutrinos may also produce a signature in cosmic ray air shower detectors.
The Pierre Auger Observatory has reported an upper limit on the tau
neutrino flux from a search for horizontal showers 
from tau lepton decays induced by Earth-skimming cosmogenic
neutrinos. However, the energy threshold of this search is 200 PeV,
 much higher than the energy of the astrophysical neutrinos observed by IceCube~\cite{Aab:2015kma}.

This paper describes a new search method for double bangs whose two
cascades may not be separately reconstructed, 
but which appear as a two-peaked or ``double pulse'' waveform in one or more IceCube sensors.

\section{The IceCube Detector}


The IceCube detector~\cite{achterberg2006first} consists of 86
vertical cables, called
strings, deployed in the ice near the geographic South Pole. Each string contains 60 Digital Optical Modules (DOMs). A DOM consists
of a 10 inch photomultiplier tube (PMT), digitizing electronics, and
LED flashers for calibration~\cite{Abbasi2010139}. 
The digitized PMT signal is called a waveform.
The DOM utilizes two digitizers: the Analog Transient Waveform
Digitizer (ATWD) which digitizes at 3.3 ns per sample for 128 samples, 
and the fast Analog to Digital Converter (fADC) which
digitizes continuously at 25 ns per sample, with a record length set at 256 samples. 
The ATWD output is separated into 3 different gain channels (x16, x2,
x0.25) to cover the dynamic range of the PMT, which has a linear
response (within 10\%) up to
currents of 400 photoelectrons (PE) per 15~ns~\cite{Abbasi2009294}. 
When a PMT receives a signal above a threshold of 0.25 PE, this is
called a hit. The x16 gain channel is captured first, with the
x2 and x0.25 channels captured if the next lowest gain channel exceeds
768 ADC counts in any sample. A local coincidence hit (LC) occurs if a pair of
nearest or next-to-nearest neighbor DOMs on the same string are hit within 1
microsecond. For LC hits, the complete ATWD and fADC waveforms will be sent
to the surface. 
The primary IceCube trigger keeps all DOM hits if 8 or more LC hits
occur anywhere in the detector within a 5 microsecond window; such a collection of
DOM hits is called an event.

IceCube employs a number of filtering algorithms in order to reduce the
data volume for transmission to the Northern hemisphere. 
The analysis described in this paper uses the ``Extremely High Energy'' (EHE) filtering
algorithm, which keeps all events that deposit more than 1000 PE in the detector.

IceCube was fully built as of December 2010. This analysis uses 914.1
days of data from the full detector between May 13, 2011 and May 6, 2014. 
The data were kept in the analysis chain only when all IceCube strings were
operating and no {\it in-situ} calibration light sources were in use.

Background and signal passing rates were computed using Monte Carlo
simulation. The CORSIKA~\cite{Heck:1998vt} simulation package is used to
generate cosmic ray induced muons. Astrophysical and atmospheric
neutrinos are simulated using an adapted version of the Monte Carlo
generator ANIS~\cite{gazizov2005anis}. 
Photon propagation through the ice is simulated as described in~\cite{aartsen2013measurement}. 
PMT response and digitization electronics are fully simulated, which
is particularly important for this analysis. 

In addition to the simulation, 10\% of the data were used to
develop cuts and estimate cosmic ray muon background rates, 
with the rest of the data not used until the cuts were finalized.

\section{Search for Tau Neutrinos}

\subsection{Double Pulse Algorithm}

\begin{figure}
 \includegraphics[width=0.5\textwidth]{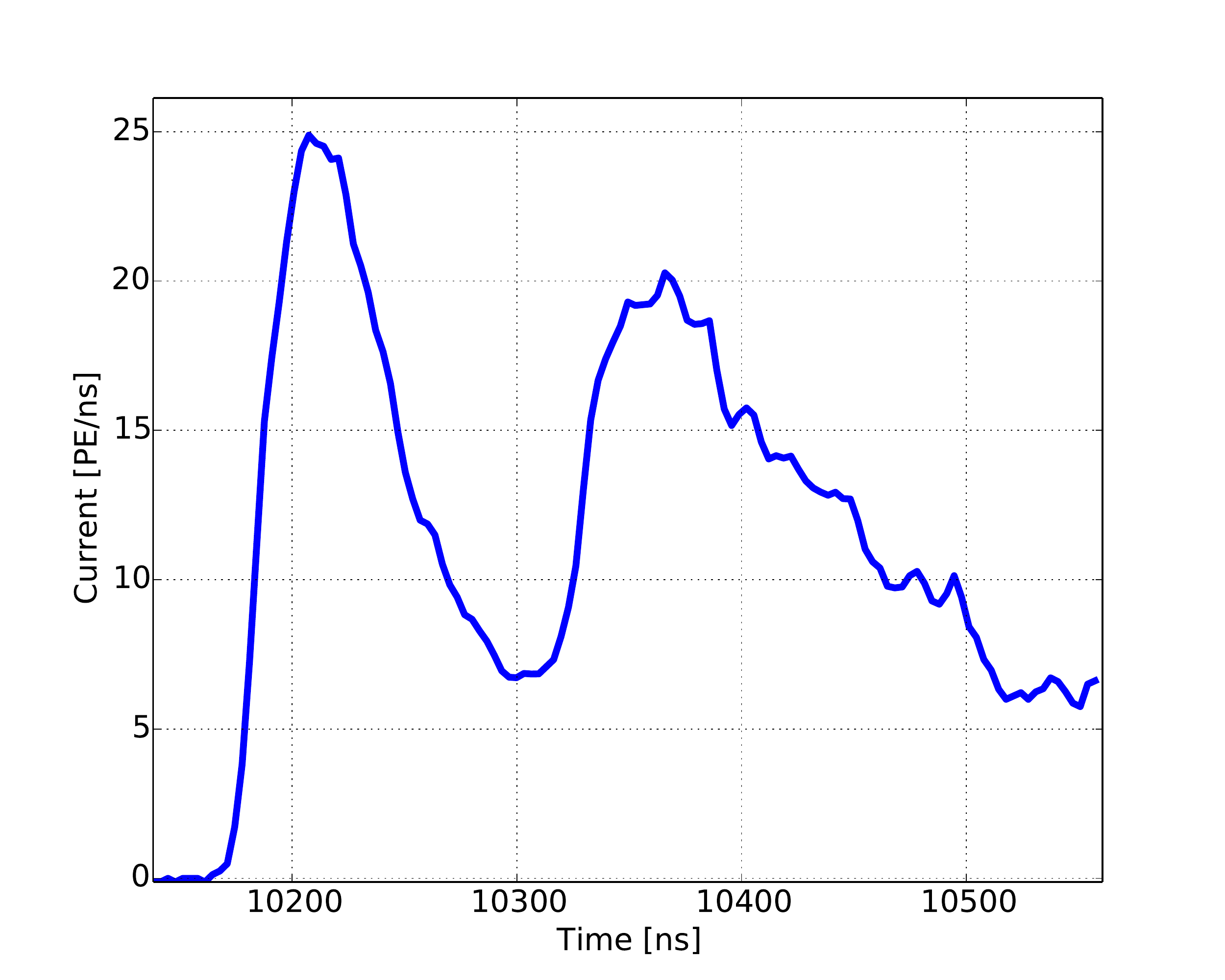}
 \caption{A simulated double pulse waveform obtained in one DOM from a $\nu_{\tau}$ CC
   event. The primary neutrino energy for this event is 2.4 PeV, about
 75\% of this energy transfers to the outgoing $\tau^{-}$ lepton, which
 travels 40 meters before decay. In this event, a total of 34 double pulse waveforms
 were produced from adjacent DOMs on neighboring strings near the
 event interaction vertices. The distances from the CC vertex and the
 $\tau^{-}$ decay vertex to the DOM that produced this double pulse
 waveform are 76 m and 75 m, respectively. Time = 0 corresponds to the
 beginning of the event readout window, which begins 10 microseconds
 before the event trigger launches. }
 \label{TauDP_wf}
\end{figure}


The goal of the double pulse algorithm (DPA) is to identify double
pulse waveforms that are consistent with $\nu_{\tau}$ CC interaction
signatures in IceCube, while rejecting waveforms with features that
are consistent with late scattered photons from single cascade
events from NC and $\nu_e$ CC interactions. There are two additional types of background events which could produce
substantial double pulse waveforms: 
(1) high energy single muons and/or muon bundles induced by cosmic
rays interacting with the atmosphere;  
(2) $\nu_{\mu}$ CC
interactions in IceCube which produce energetic muons.
For double pulse waveforms caused by energetic atmospheric muons, 
the first pulse is usually from a combination of Cherenkov light
emissions and coincident stochastic energy loss, 
and the second pulse is from TeV-scale stochastic
energy losses tens of meters away from the DOM. 
For double pulse waveforms from astrophysical $\nu_{\mu}$ CC events, 
the first pulse is from energy deposition of the CC hadronic
interaction vertex, while the second pulse is from a coincident
stochastic energy loss of the energetic outgoing muon. 
Since the double pulse waveforms from $\nu_{\tau}$ CC events, energetic atmospheric
muons and astrophysical $\nu_{\mu}$ CC events are not distinguishable from
one another as they are caused by the same mechanism of two
substantial energy depositions near certain DOMs, 
we do not remove these events with the DPA. 
They are to be removed later by comparing their overall topologies and
timing profiles.
The potential impact of instrumental backgrounds on the DPA was found
to be negligible. Afterpulses in the PMT waveforms, caused by
ionization of residual gases by electrons accelerated in the space
between dynodes, usually occur from 500 ns to microseconds later than
the primary pulse, and therefore do not appear as double pulses in a
single waveform. Late pulses, caused by photoelectrons backscattered
from the first dynode, occur on a time scale of 60 ns later than the
primary pulse, but usually have a low amplitude and do not trigger the
DPA \cite{Abbasi2010139}.

The DPA uses the positive and negative first derivatives
of a waveform to determine rising and trailing edges. A double pulse is
defined as a rising edge, followed by a trailing edge, 
followed by another rising edge. Waveforms from the ATWD digitizer in the lowest gain
channel available are used since higher gain channels are usually
saturated for high amplitude waveforms. The fADC waveforms are not used
 since they do not have multiple gain channels available and since
 their coarser timing causes double pulse features to be blended
 together or saturated. 
The DPA uses 7 configurable parameters to characterize a double pulse
waveform:

\begin{itemize}
\item Since signal waveforms are from
bright events close to a DOM, the DPA is only run on ATWD waveforms
that have integrated charge greater than $q_1$ = 432 PE. 

\item The beginning of the waveform is determined by a
 sliding time window of 3.3 ns which searches for
 a monotonic increase in the waveform amplitude within a time span of
 3.3 $\times$ 6 = 19.8 ns.

\item Once the beginning of the waveform is identified, the waveform is
divided into segments of 4 ATWD bins (13.2 ns) and the first derivative of the waveform is computed in
each segment. 

\item If the first derivative is positive in $n_1$ = 2 consecutive segments, this
is considered the rising edge of the first pulse. 
When the subsequent derivative is negative for $n_2$ = 2 consecutive
segments, this is considered the trailing edge of the first pulse. 
The rising edge of the first pulse is required to have an integrated
charge of at least $q_2$ = 23 PE, and the integrated charge of the trailing
edge is required to be at least $q_3$ = 39 PE. 
The integrated charge sums up all the charge corresponding to the
entire rising or trailing edges, 
which usually last longer than two segments (26.4 ns) for a large pulse. 

\item The second pulse rising edge is defined when the derivative after the
trailing edge of the first pulse is positive again for $n_3$ = 3
consecutive segments. This requirement is due to the fact that the
light in the second pulse is often more scattered and therefore has 
a less steep rising edge than the first pulse. The second pulse
trailing edge is often outside the ATWD window, 
and hence is not included in the calculation. The rising edge of the second pulse is
required to have an integrated charge of at least $q_4$ = 42 PE.  

\end{itemize}

The configurable DPA parameters $n_i$ and $q_i$ were tested and optimized using a
variety of IceCube event waveforms including simulated neutrinos of all
flavors, simulated atmospheric muons and data from {\it in-situ}
laser calibration devices. 
An example of a simulated double pulse waveform from a
$\nu_{\tau}$ CC event is shown in Figure~\ref{TauDP_wf}. The production
of double pulse waveforms depends largely on the distances between event interaction
vertices and nearby DOMs. The median distance  
between the $\nu_{\tau}$ CC (tau lepton decay) vertices and the double
pulse DOMs is 49 (44) meters.

\begin{figure}
  \includegraphics[width=0.5\textwidth]{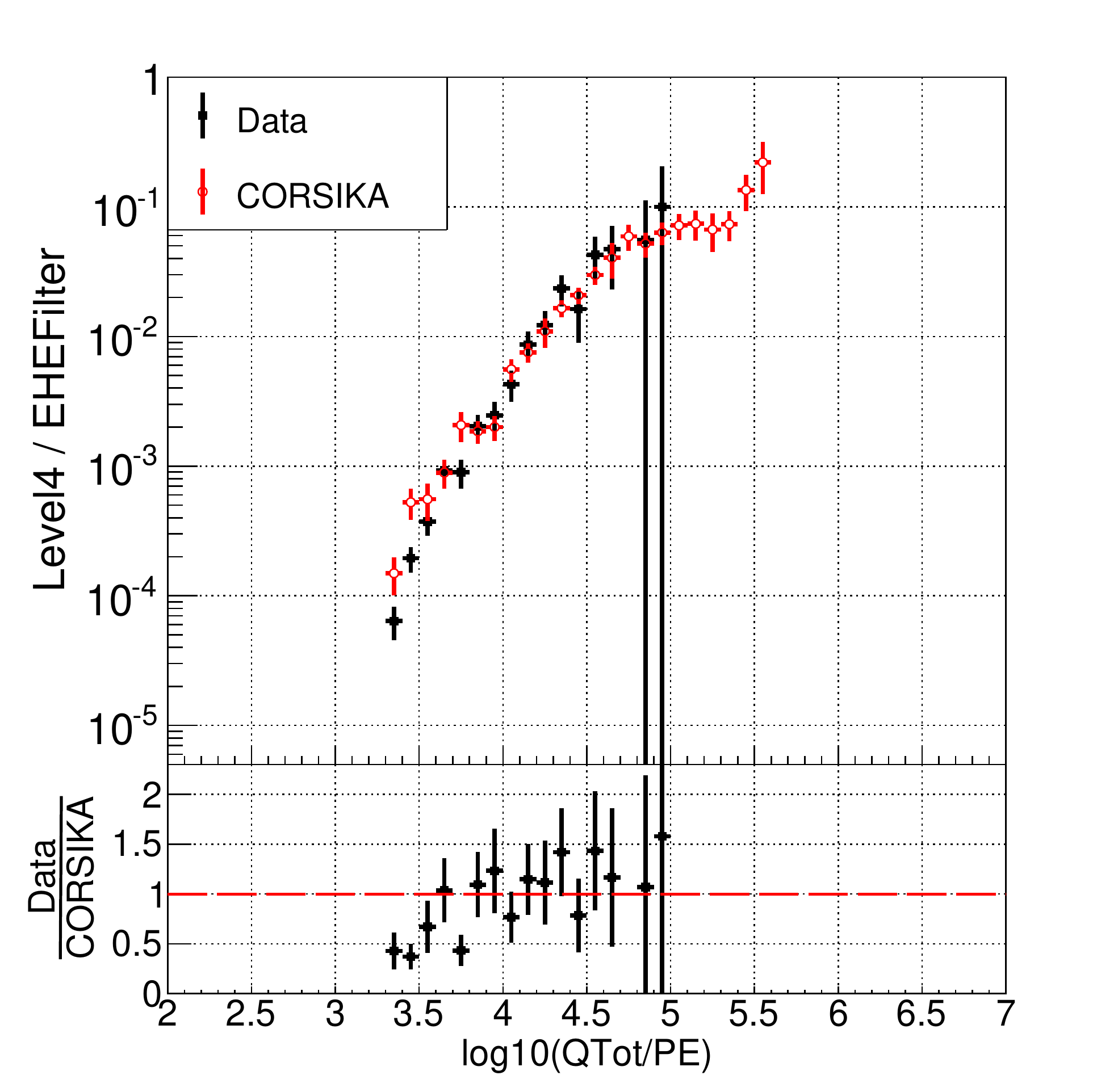}
 \caption{Fraction of events that pass the charge cut of log$_{10}$(QTot)$>$3.3 and have at
   least one double pulse waveform as a function of total deposited
   charge. The higher the charge in the event, the more likely it is
   to contain at least one double pulse waveform. Muons with lower
   deposited charge that might evade containment cuts are less likely to 
produce double pulse waveforms.} 
 \label{Dp_prob}
\end{figure}

The fraction of events that pass a charge cut of log$_{10}$(QTot)$>$3.3 and have at least one double pulse waveform is 
shown in Figure~\ref{Dp_prob} as a function of
deposited charge, for both the 10\% data sample and atmospheric muon
simulation. Near the charge cut threshold, 
fewer than 1 in 1000 events will include a double pulse waveform. 

\subsection{Event Selection} \label{evt_select}

The event selection process was carried out at three cut levels, driven by the specific
goal of background rejection at each level. To conform with standard
IceCube usage, the cuts for this analysis are numbered beginning with
Level 4. The cut levels are summarized as follows: 

\begin{figure*}[!tp]
  \centerline{\includegraphics[width=0.5\textwidth]{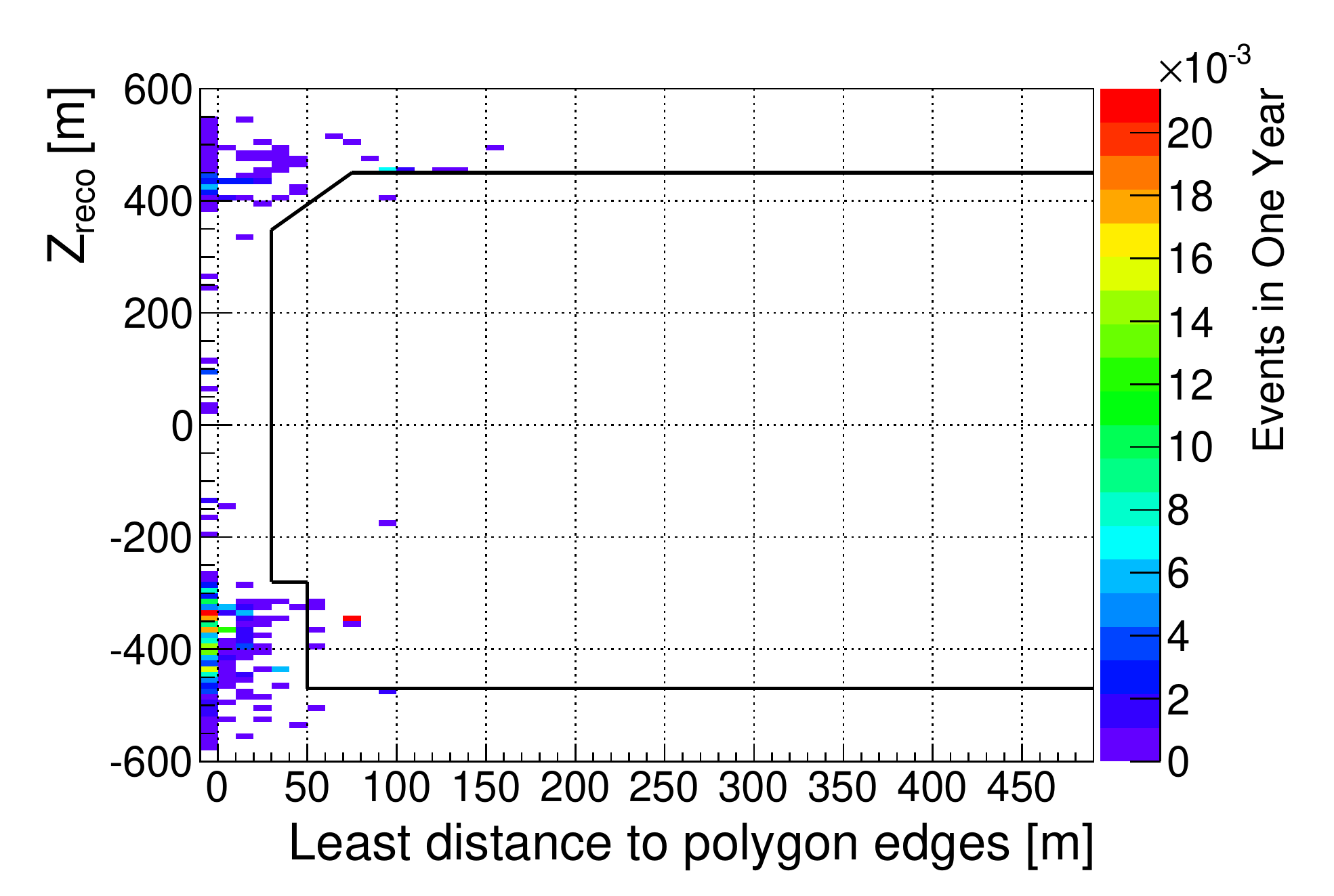}\label{fig2}
             \hfil
             \includegraphics[width=0.5\textwidth]{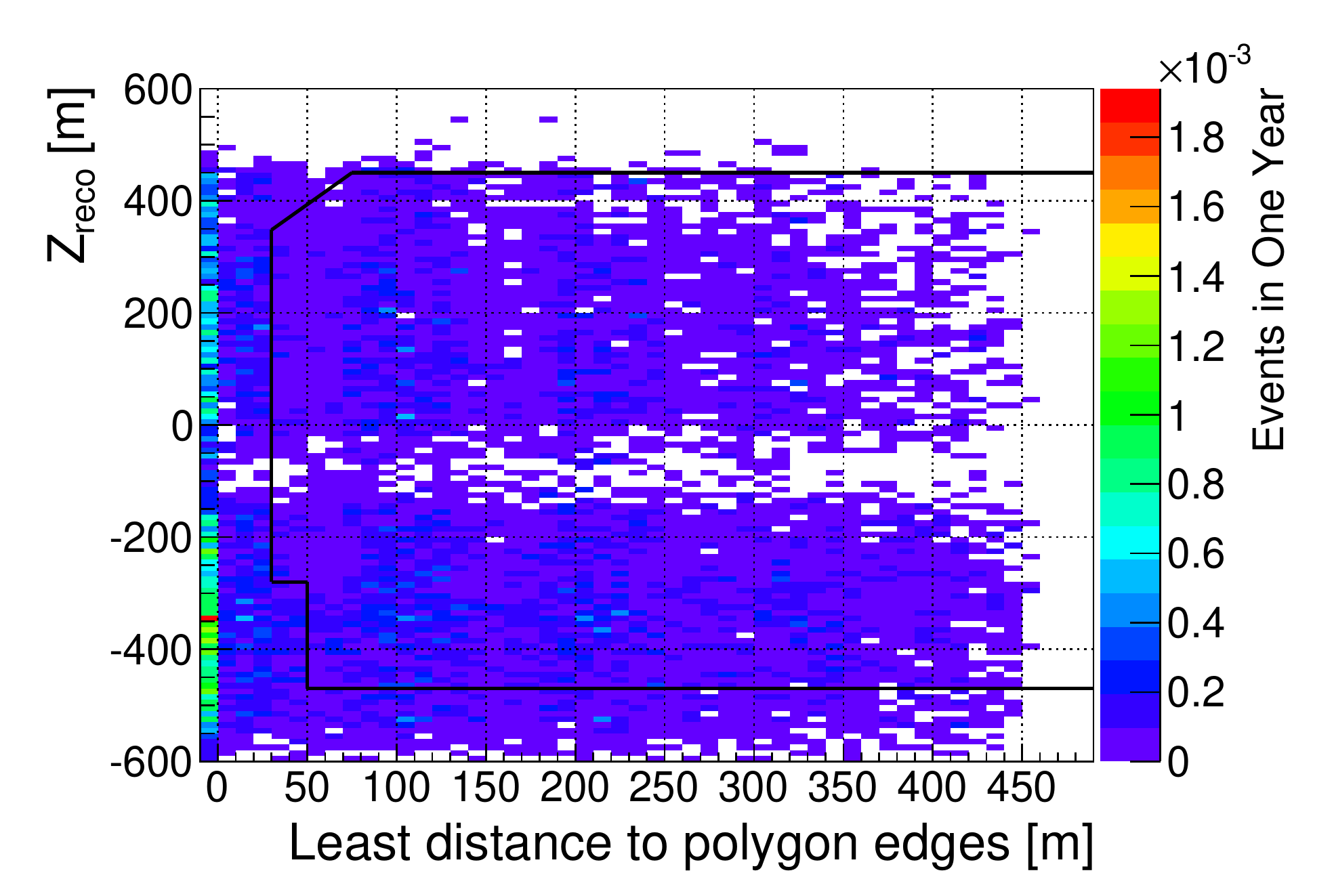} \label{fig3}
            }
  \caption{Events per year that pass the level 6 containment cut on depth and
    distance to the detector edge (solid line) for simulated
    atmospheric muons (left) and astrophysical $\nu_{\tau}$ (right).
           }
  \label{containment}
\end{figure*}

Level 4: Events are required to have at least one waveform which
passes the DPA.  An additional event-wise charge cut of
log$_{10}$(QTot)$>$3.3 is also required to enrich the sample with high
energy events.

Level 5: At this stage, we remove track-like double pulse events which
are predominantly due to atmospheric muons. Following the Level 4 cut, each event is reconstructed
using a maximum likelihood method based on a hypothesis of an infinite track and a hypothesis of a
point-like cascade. These reconstructions only make use of the timing information
for the earliest photon arriving at the DOMs and hence are
computationally efficient. The reduced log likelihood ratio between the two hypotheses
L$_{R}$=log(L$_{\text{cascade}}$/L$_{\text{track}}$) is required to be negative, 
indicating the event topology is more cascade-like than track-like. 
This cut eliminates most down-going energetic muons and muon bundles. 
To further veto down-going muons, the first hit in the
event is required to be below the top 40 meters of the instrumented
volume. CORSIKA simulation predicts that 3.5$\pm$3.4 atmospheric muons survive to Level 5 in 914.1 days.

Level 6: At this stage we eliminate cosmic ray induced muons which pass near the edges of the
detector and hence appear cascade-like. An additional reconstruction algorithm is performed on all events
which pass the preceding cuts, using full charge and time
information, which is more computationally expensive. A boundary is defined by the surface connecting the position
of the outermost layer of strings in the detector. 
The containment criterion requires that the reconstructed vertex be
inside the instrumented volume and a given distance away from the
boundary. The distance from the boundary depends on depth, with
stricter containment required at the top and bottom of the
detector. The containment is illustrated in Figure~\ref{containment},
which shows the distribution of event vertices with respect to the boundary
for signal and for atmospheric muon background. 
Due to the scarcity of events in the atmospheric muon simulation at high energies,
the double pulse criterion is removed from this plot, 
with all other cuts kept. The very few atmospheric muon events which
survive the containment cut are close to the charge cut threshold, 
and as shown in Figure~\ref{Dp_prob}, such events have a lower probability of producing a double pulse waveform.

\begin{figure}
 \includegraphics[width=0.5\textwidth]{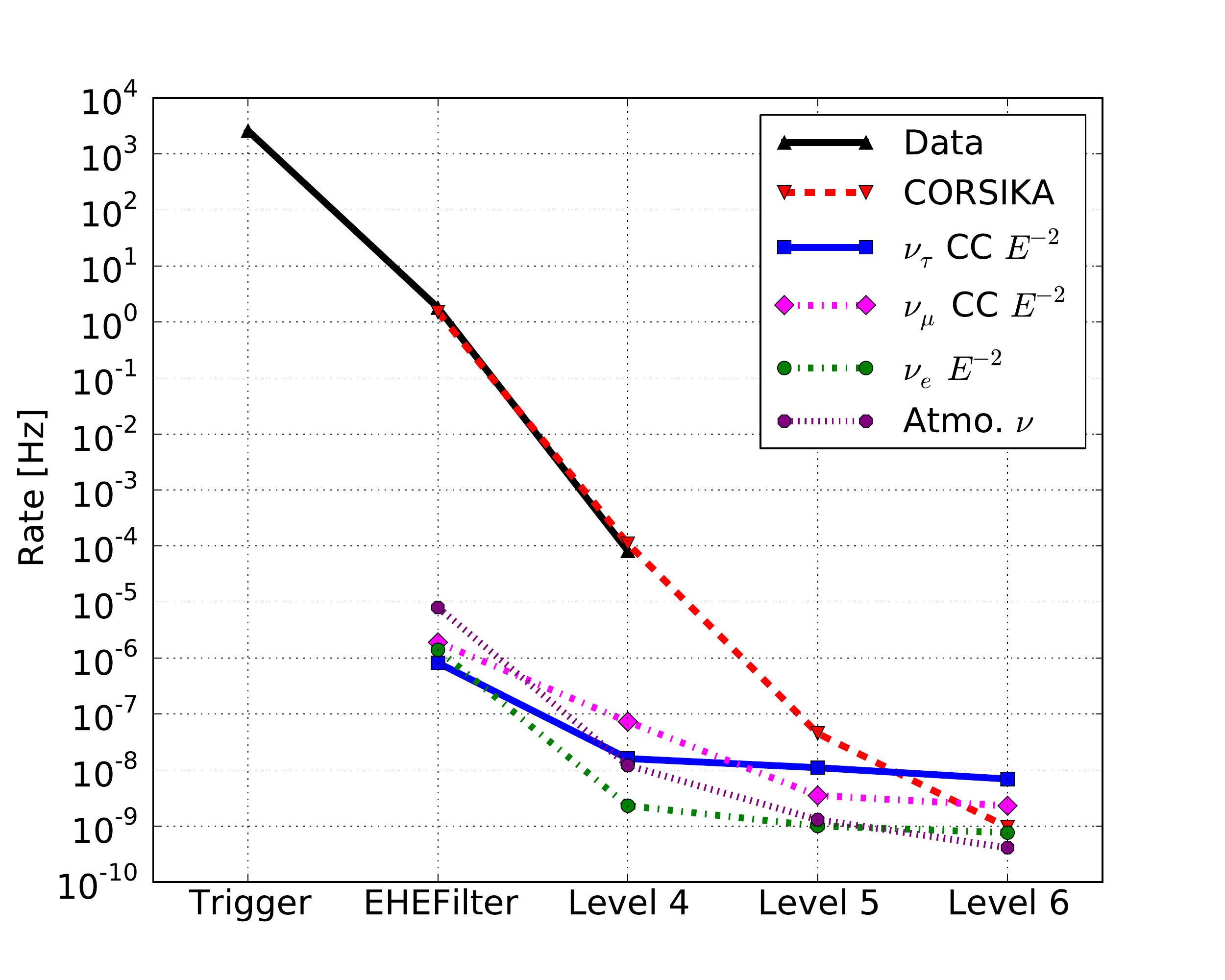}
 \caption{Passing rate for signal (astrophysical $\nu_{\tau}$ CC events in
  solid blue) and backgrounds (astrophysical $\nu_{\mu}$ in dot-dashed
   magenta, astrophysical $\nu_e$ in dot-dashed green, atmospheric
   neutrinos in dotted purple, and atmospheric muons or CORSIKA  in dashed red)
   as a function of cut level. Data shown here in solid black is 10\% of
   the total data sample.}
 \label{CutEff}
 \end{figure}

At the final cut level, the predicted rates from all sources in three
years of data are summarized in Table~\ref{evt_rates}. 
The assumed astrophysical flux is based on the diffuse flux measured
by IceCube at the level of E$^2$$\Phi_{\nu}$ = 1.0 x 10$^{-8}$ GeV cm$^{-2}$ s$^{-1}$
sr$^{-1}$ per flavor~\cite{aartsen2014observation}, and 90\% of the
predicted $\nu_{\tau}$ CC events are between 214 TeV and 72 PeV. 
A softer neutrino energy spectrum of E$^{-2.5}$ \cite{Aartsen:2015knd} reduces the
expected number of $\nu_{\tau}$ events and the dominant background of $\nu_{\mu}$ CC
events by 36\% and 57\% respectively. 
The atmospheric neutrino rate prediction includes both neutrinos from $\pi/K$
decay~\cite{PhysRevD.75.043006} and neutrinos from charmed meson
decay~\cite{PhysRevD.78.043005}. The primary cosmic ray spectrum used
to predict atmospheric neutrino rates is corrected for air shower
measurements in the knee region of several
PeV~\cite{gaisser2012spectrum}.

\begin{table}
  \caption{Predicted event rates from all sources at the final cut
    level. Errors are statistical only. \label{evt_rates}}
  \begin{ruledtabular}
    \begin{tabular}{| c | c |}
      Data samples & Events in 914.1 days (final cut)\\ 
      \hline
      Astrophysical $\nu_{\tau}$ CC & (5.4 $\pm$ 0.1) $\cdot$ 10$^{-1}$\\
      Astrophysical $\nu_{\mu}$ CC & (1.8 $\pm$ 0.1) $\cdot$ 10$^{-1}$ \\
      Astrophysical $\nu_e$ & (6.0 $\pm$ 1.7) $\cdot$ 10$^{-2}$  \\
      Atmospheric $\nu$ & (3.2 $\pm$ 1.4)  $\cdot$ 10$^{-2}$  \\
      Atmospheric muons~~~~ & (7.5 $\pm$ 5.8)  $\cdot$ 10$^{-2}$  \\
    \end{tabular}
  \end{ruledtabular}
\end{table}

\begin{figure}
  \includegraphics[width=0.5\textwidth]{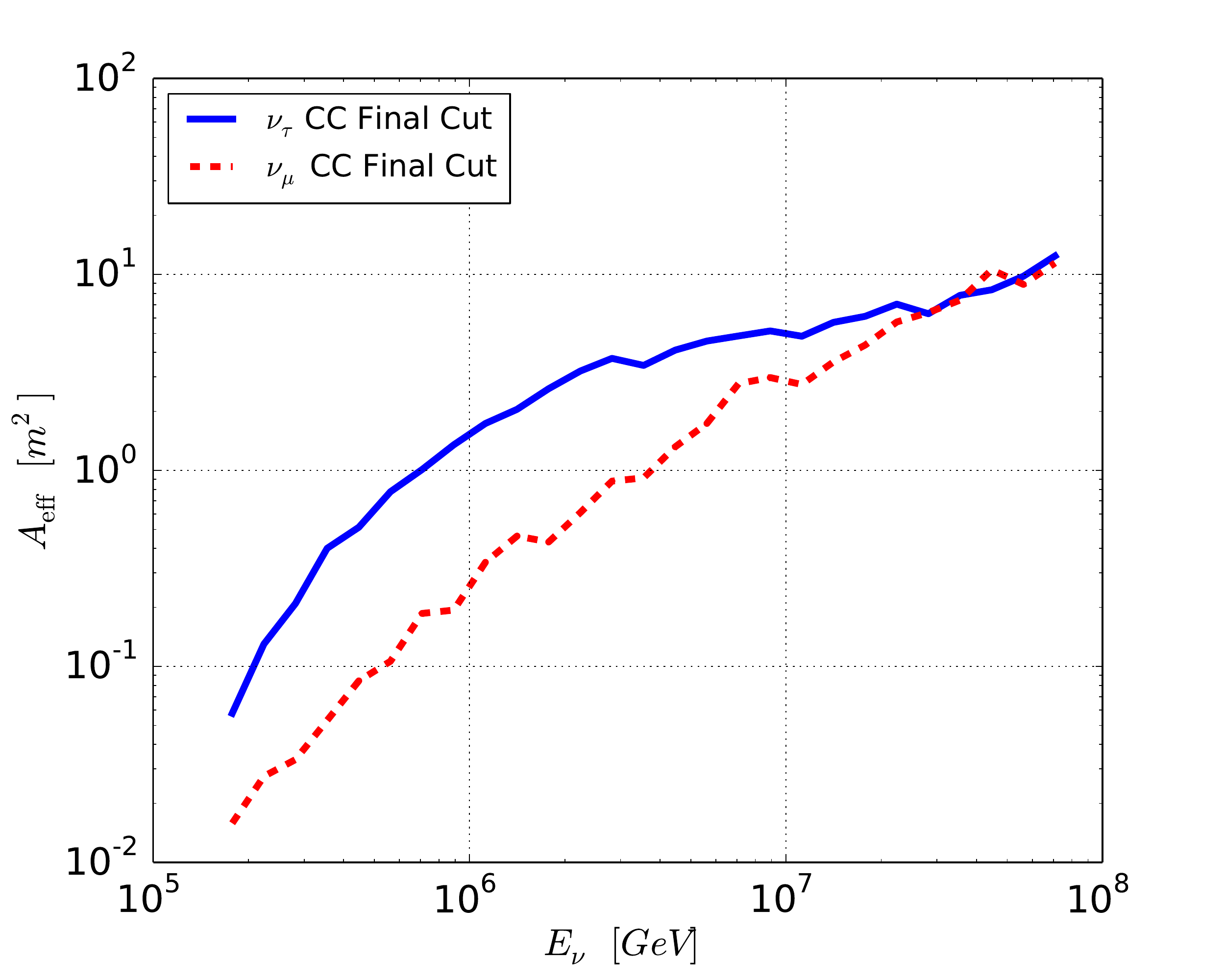}
  \caption{Effective areas at final cut level as a function of primary
    neutrino energy. Only the middle 90\% $\nu_{\tau}$ energy range
    (214 TeV - 72 PeV) is plotted. The dominant background for this
    analysis is due to astrophysical $\nu_{\mu}$ CC events, so only $\nu_{\tau}$
    CC (solid blue) and $\nu_{\mu}$ CC (dashed red) effective areas are shown. 
The plot demonstrates that the optimal energy window for the
astrophysical $\nu_{\tau}$ search using the double pulse waveform
approach is from O(100) TeV to O(10) PeV. In particular, around
PeV energies, effective areas for $\nu_{\tau}$ CC events are about an order of magnitude higher
than those for $\nu_{\mu}$ CC events. Effective areas for $\nu_e$ (not shown)
are 1-2
orders of magnitude below the effective areas for  $\nu_{\tau}$ CC,
except at the Glashow resonance energy of 6.3~PeV~\cite{Glashow:1960zz}.
}
  \label{Aeff}
\end{figure}

Figure~\ref{CutEff} summarizes the passing rate of signal and background events at each cut level.
At the final cut level, astrophysical $\nu_{\tau}$ events have the highest passing rate of any
source, and the dominant background is astrophysical $\nu_{\mu}$ CC
events. The effective areas for $\nu_{\tau}$ CC and $\nu_{\mu}$ CC
events at the final cut level is shown in Figure~\ref{Aeff}. An optimal
energy window for the astrophysical $\nu_{\tau}$ search in IceCube using this double pulse
method is around the PeV region, where the effective areas for
$\nu_{\tau}$ CC events are nearly an order of magnitude higher than
that of $\nu_{\mu}$ CC events.
It is planned that events found at final cut level will be further
investigated with segmented energy loss reconstruction algorithms
\cite{1748-0221-9-03-P03009} to acquire their energy loss profile and directionality. Event
probabilities of $\nu_{\tau}$-like or not will also be computed
based on likelihood methods.   

\section{Systematic Uncertainties}

The sources of systematic uncertainty considered in this analysis are neutrino cross sections,
anisotropy in the optical scattering in ice, uncertainty in the optical scattering and absorption lengths in ice, 
and DOM efficiency. The main sources of systematic uncertainty in the
signal are summarized in Table~\ref{sys}. 

The neutrino cross sections used in this analysis are from the CTEQ5
model \cite{lai2000global}. The CSMS model~\cite{CooperSarkar:2011pa},
which has updated parton distribution functions, predicts $\sim$ 5\% fewer events compared to the CTEQ5 model. 

An earlier study in IceCube attempting to reconstruct the double deposition of
energy from a $\nu_{\tau}$ CC event has found that the recently
identified anisotropy in the optical scattering in ice
\cite{aartsen2013icecube} would modify
the number of expected photons in some 
DOMs and hence could mimic a double cascade feature in the reconstructed
energy segments \cite{hallen_thesis}. A
study based on simulations with and without this anisotropy found a
7\% lower signal event rate prediction for the double pulse analysis when
anisotropy was included. The effect is small at the waveform level due to the fact that the
double pulse events usually occur within tens of meters of a DOM,
which is within 1-2 scattering lengths in the ice.

The optical scattering length and absorption length were varied
according to the uncertainty in the value of these
parameters~\cite{aartsen2013measurement}. Increasing the absorption by the allowed uncertainty
decreases the signal event rate by 4.9\%, and decreasing the absorption and
scattering increases the signal event rate by 8.1\%.

Since the $\nu_{\tau}$ double pulse events are very bright, uncertainty in the
DOM efficiency does not play an important role. Simulation with the
DOM efficiency set at +10\% and -10\% of the nominal values yielded a
decrease of 1.6\% in the signal event rate when decreasing the efficiency,
and an increase of 6.7\% in the signal event rate when increasing the
efficiency. 

Adding the various errors in quadrature, the total systematic
uncertainty in the signal is about $\pm 10$\%.

The uncertainty in the atmospheric muon and neutrino background is
dominated by statistical error, due to the fact that few simulated
background events pass the cuts. The largest source of systematic
error is uncertainty in the cosmic ray flux at high energies which
contributes +30\%/-50\% uncertainty to the atmospheric muon flux and
$\pm 30$\% uncertainty to the atmospheric neutrino flux~\cite{Aartsen:2013dsm}.

\begin{table}
  \caption{Source of systematic uncertainty in the signal. \label{sys}}
  \begin{ruledtabular}
    \begin{tabular}{| c | c |}
      Neutrino cross sections~~~~ & -5\% \\ [1ex]
      \hline
      Anisotropy in the optical scattering in ice & -7\% \\ [1ex]
        \hline
      Optical scattering and absorption lengths in ice &
      $_{-4.9\%}^{+8.1\%}$ \\ [1ex]
       \hline
      DOM efficiency & $_{-1.6\%}^{+6.7\%}$ \\ [1ex]
      \hline
      Total & $_{-10.0\%}^{+10.5\%}$
    \end{tabular}
  \end{ruledtabular}
\end{table}

\begin{figure*}[!tp]
  \centerline{\includegraphics[width=0.4\textwidth]{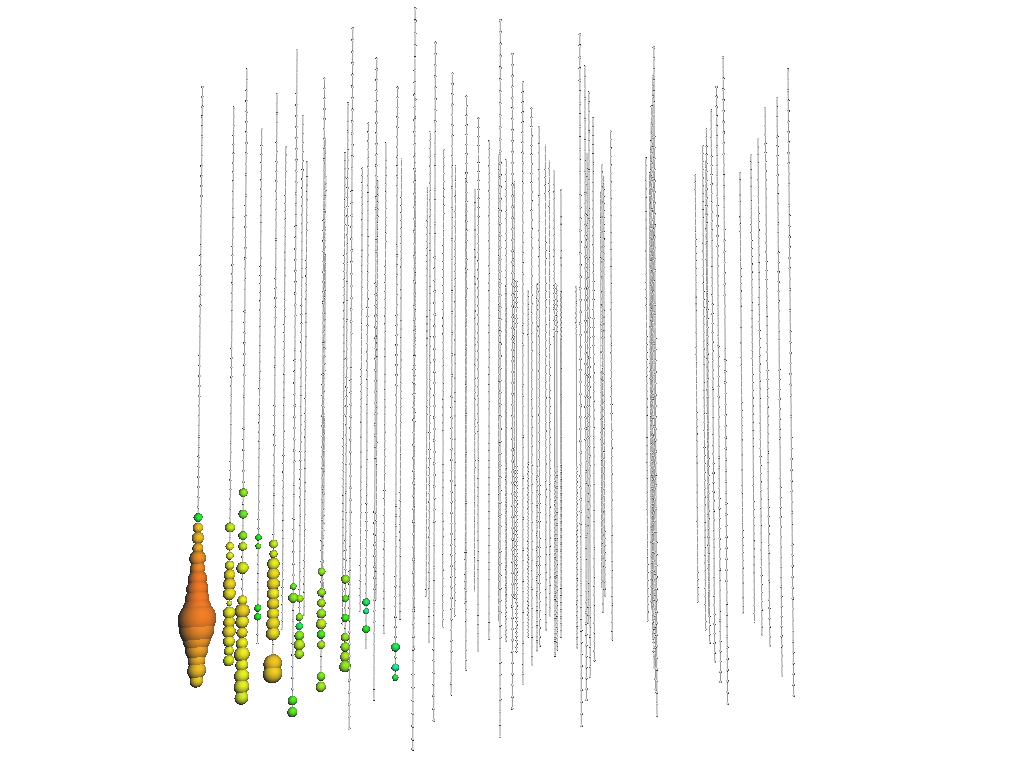}\label{fig2a}
             \hfil
             \includegraphics[width=0.4\textwidth]{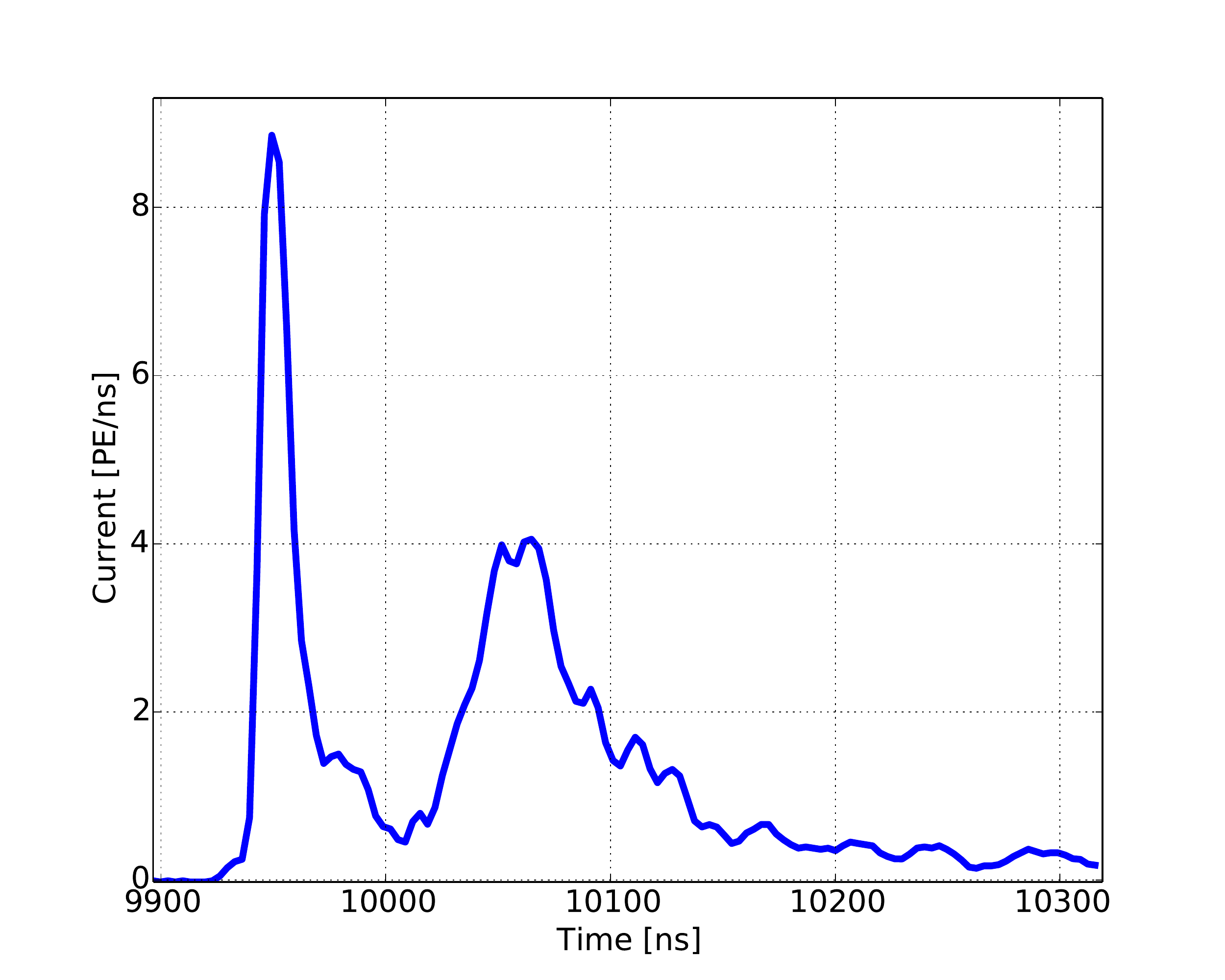} \label{fig3a}
            }
  \caption{Event 1 before level 6 containment cut with its corresponding double
    pulse waveform. This event occurred on May 30, 2011. The colored
    spheres indicate hit DOMs, with size indicating the amount of
    charge deposited on the sphere and color indicating time: red is
    earlier, blue is later.
           }
  \label{double_fig1}
\end{figure*}

\begin{figure*}[!tp]
  \centerline{\includegraphics[width=0.4\textwidth]{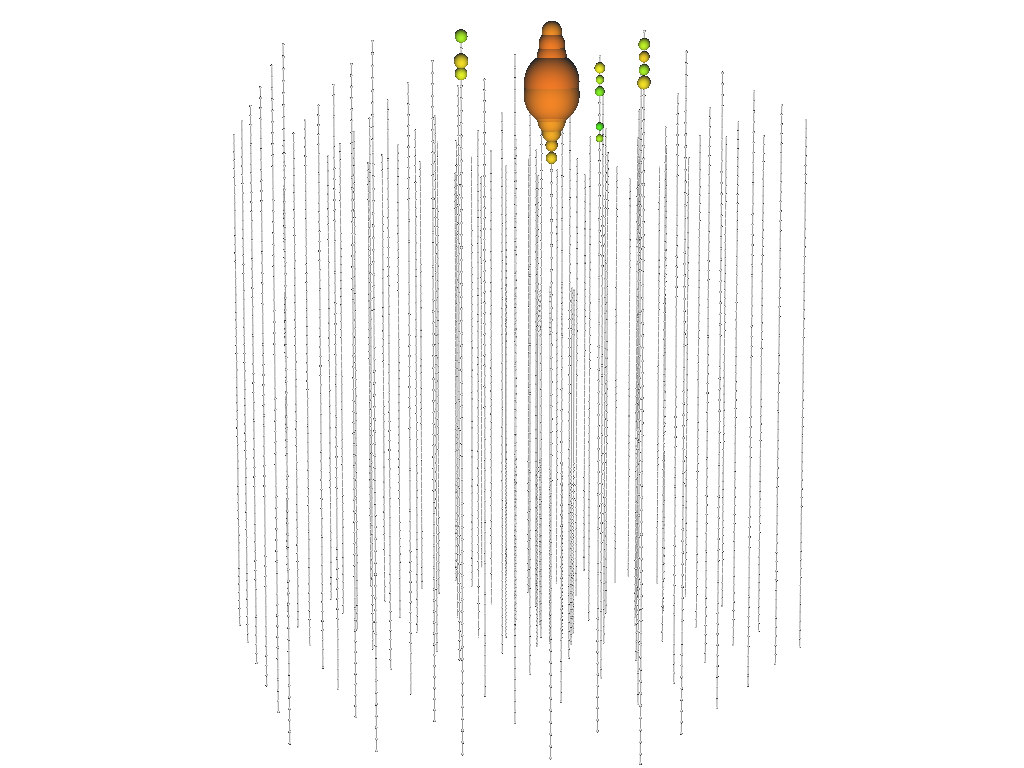}\label{fig2b}
             \hfil
             \includegraphics[width=0.4\textwidth]{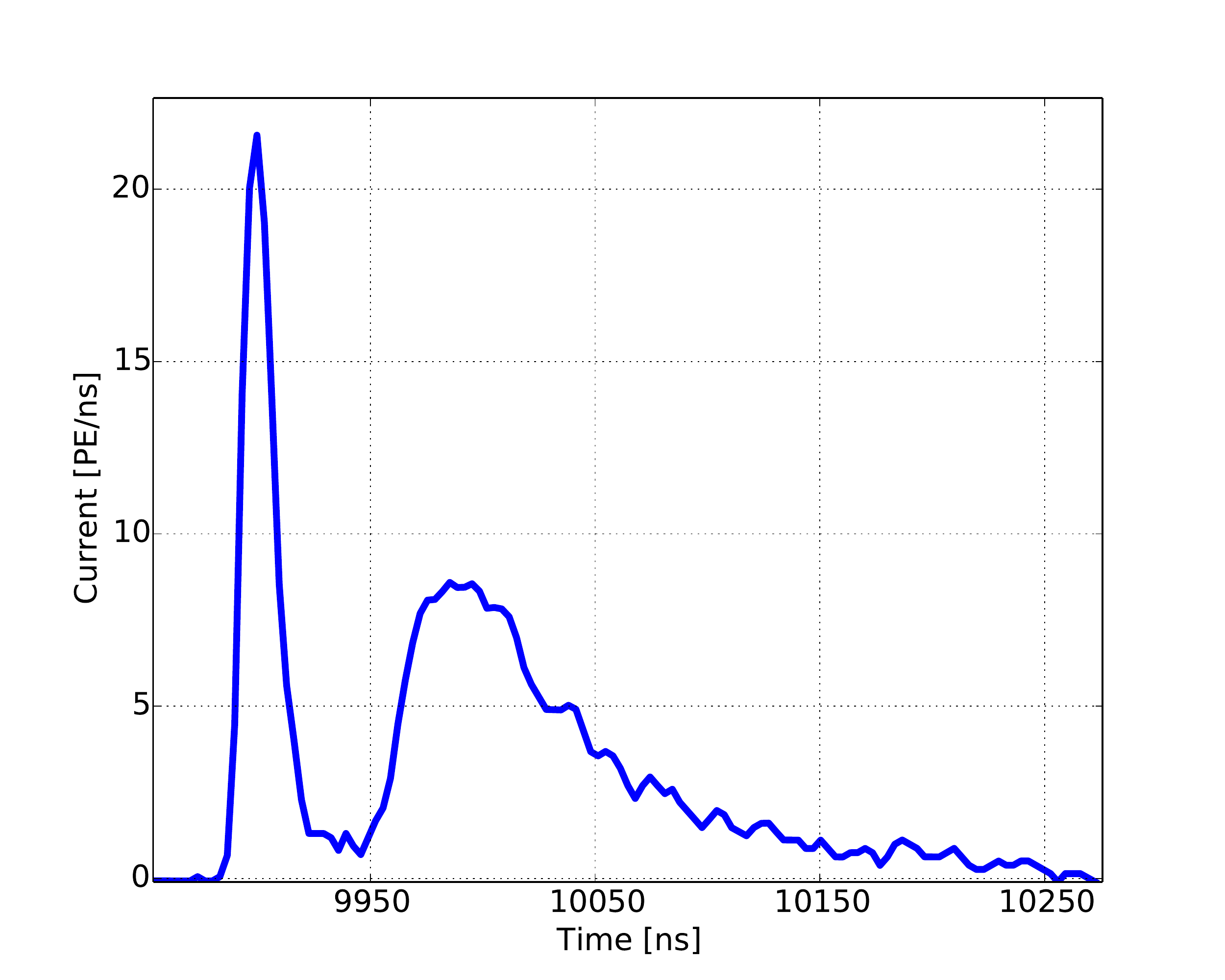} \label{fig3b}
            }
  \caption{Event 2 before level 6 containment cut with its corresponding double
    pulse waveform. This event occurred on November 27, 2011.
           }
  \label{double_fig2}
\end{figure*}

\begin{figure*}[!tp]
  \centerline{\includegraphics[width=0.4\textwidth]{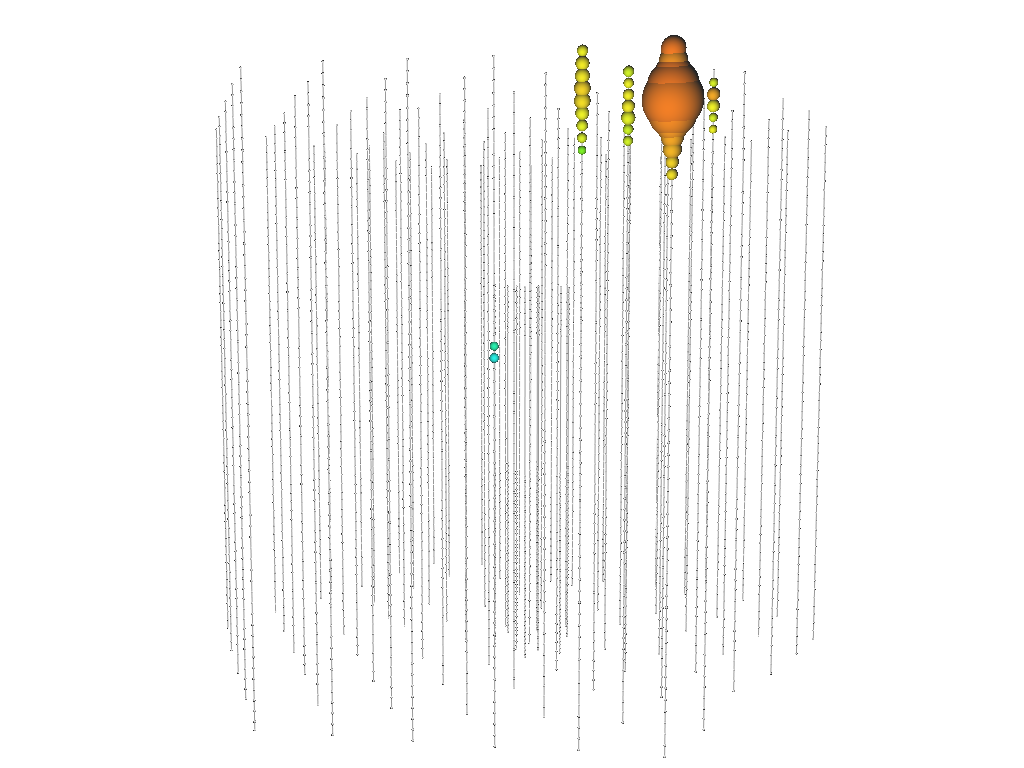}\label{fig2c}
             \hfil
             \includegraphics[width=0.4\textwidth]{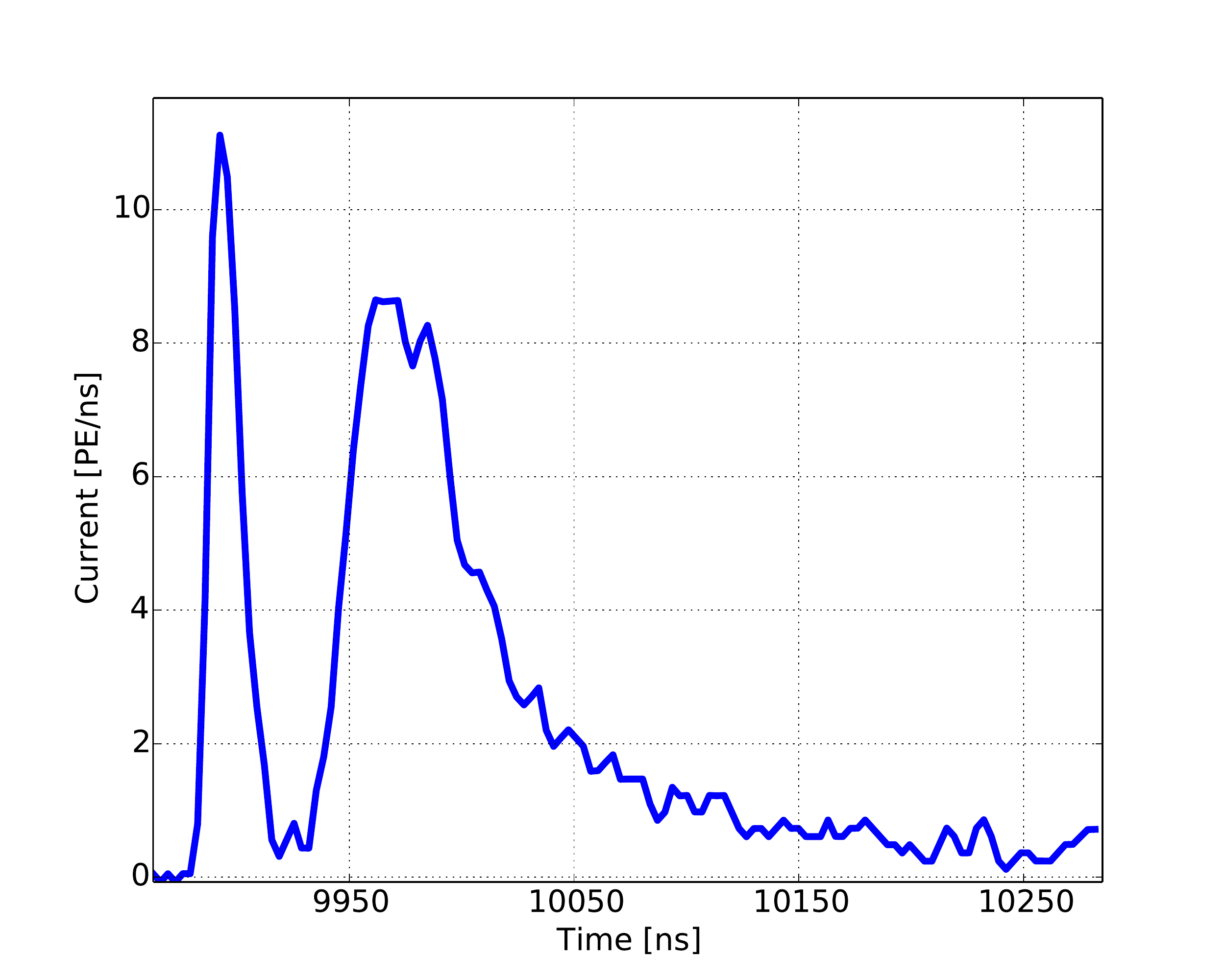} \label{fig3c}
            }
  \caption{Event 3 before level 6 containment cut with its corresponding double
    pulse waveform. This event occurred on August 28, 2012.
           }
  \label{double_fig3}
\end{figure*}

\section{Results}

\begin{figure}
 \includegraphics[width=0.5\textwidth]{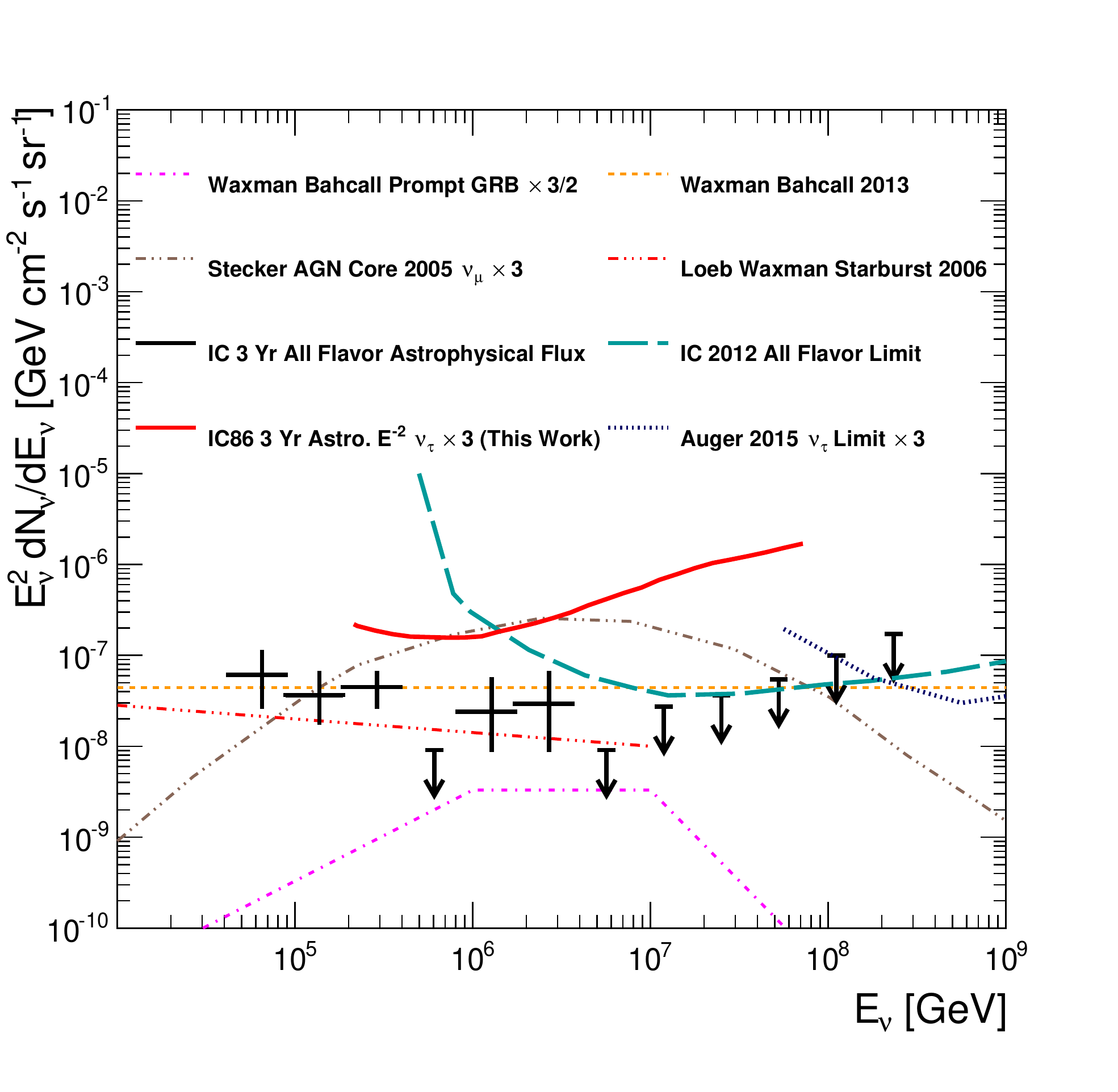}
 \caption{Neutrino flux upper limits and models as a function of
   the primary neutrino energy. The thick red curve is the $\nu_{\tau}$
   differential upper limit derived from this analysis, including
   systematic and statistical errors. In computing the differential
   upper limit, values of the flux limit were calculated for each energy decade with a sliding
   energy window of 0.1 decade. The thick black error bars depict the all-flavor astrophysical neutrino flux observed by IceCube
   \cite{aartsen2014observation}. The thick dashed line
   is the differential upper limit derived from a search for extremely
   high energy events which has found the first two PeV cascade events
   in IceCube \cite{Aartsen:2013bka, Aartsen:2013dsm}. The blue
   dotted line is the Auger differential upper
   limit from $\nu_{\tau}$ induced air showers
   \cite{Aab:2015kma}. The orange dashed
   line is the Waxman-Bahcall upper bound
   which uses the UHECR flux to set a bound on astrophysical neutrino
   production \cite{Waxman:2013zda}. The dash-dotted line (magenta)
   represents the prompt neutrino flux predicted
 from GRBs; prompt in this context means in time with the gamma rays
 \cite{Waxman:1997ti}. The dash-dot-dot line (grey) indicates the neutrino flux predicted from the
 cores of active galaxies \cite{Stecker:2005hn}. The thin dash-triple-dot line
 (red) shows the neutrino flux
 predicted from starburst galaxies, which are rich in supernovae \cite{Loeb:2006tw}.}
 \label{UpperLimitPlot}
 \end{figure}

Zero events were found after all cuts were applied. At level 5, before the containment cut,
three events were found which each have one double pulse waveform, all of which
occurred on strings at the edge of IceCube. These events are
consistent with atmospheric muons interacting near the edge of the
detector, producing a double pulse waveform in a cascade-like event
but failing the subsequent containment cut at Level 6. The observation
of 3 events in 914.1 days of livetime matches the CORSIKA prediction at
level 5 as discussed in Section~\ref{evt_select}. The events and their
corresponding double pulse waveforms are shown in
Figures~\ref{double_fig1},~\ref{double_fig2}
and~\ref{double_fig3}.

Based on zero observed events, an integrated astrophysical $\nu_{\tau}$ flux upper limit is set to be
E$^2$$\Phi_{\nu_{\tau}}$ = 5.1 $\times$ 10$^{-8}$ GeV cm$^{-2}$ sr$^{-1}$ s$^{-1}$. 
A $\nu_{\tau}$ flux differential upper limit in the energy range of
214 TeV to 72 PeV, which contains 90\% of the predicted $\nu_{\tau}$ CC
events, is also extracted following the procedure that was
employed in deriving quasi-differential upper limits from previous
EHE cosmogenic neutrino searches in IceCube~\cite{PhysRevD.82.072003,
  PhysRevD.83.092003, Aartsen:2013dsm}. In this procedure, flux
limits were computed for each energy decade with a sliding energy window of
0.1 decade, assuming a differential neutrino flux proportional to 
$1/E^2$~\cite{PhysRevD.73.082002}. 
Since zero events were found, the 90\% C.L. event count
limit in each energy decade is 2.44 based on the Feldman-Cousins
approach~\cite{feldman1998unified}. The dominant sources of systematic error
in this analysis are independent of energy. Therefore, all the sources
of systematic and statistical error are incorporated in the limit calculation by uniform scaling of
the effective area.
The differential upper limit is plotted in Figure~\ref{UpperLimitPlot}.

\section{Conclusion}

The double pulse search method is shown to be robust, with the
observed background from cosmic ray induced muons matching
prediction. The search is more sensitive to tau neutrinos between 214
TeV and 72 PeV than to any other flavor. 
Given the astrophysical neutrino flux observed by IceCube, fewer than
one tau neutrino candidate event is expected in three years of
IceCube data, and none are observed. A differential upper limit has
been placed on the astrophysical tau neutrino flux, 
with an energy threshold three orders of magnitude lower than previous dedicated
tau neutrino searches by cosmic ray air shower detectors.  
Searches for double bang events with well separated cascades in
IceCube are underway. Future extensions of IceCube such as the proposed IceCube-Gen2
detector~\cite{Aartsen:2014njl} will 
have a factor of 5 to 10 times more sensitivity to astrophysical tau neutrinos than the current IceCube detector.

\begin{acknowledgments}

We acknowledge the support from the following agencies:
U.S. National Science Foundation-Office of Polar Programs,
U.S. National Science Foundation-Physics Division,
University of Wisconsin Alumni Research Foundation,
the Grid Laboratory Of Wisconsin (GLOW) grid infrastructure at the University of Wisconsin - Madison, the Open Science Grid (OSG) grid infrastructure;
U.S. Department of Energy, and National Energy Research Scientific Computing Center,
the Louisiana Optical Network Initiative (LONI) grid computing resources;
Natural Sciences and Engineering Research Council of Canada,
WestGrid and Compute/Calcul Canada;
Swedish Research Council,
Swedish Polar Research Secretariat,
Swedish National Infrastructure for Computing (SNIC),
and Knut and Alice Wallenberg Foundation, Sweden;
German Ministry for Education and Research (BMBF),
Deutsche Forschungsgemeinschaft (DFG),
Helmholtz Alliance for Astroparticle Physics (HAP),
Research Department of Plasmas with Complex Interactions (Bochum), Germany;
Fund for Scientific Research (FNRS-FWO),
FWO Odysseus programme,
Flanders Institute to encourage scientific and technological research in industry (IWT),
Belgian Federal Science Policy Office (Belspo);
University of Oxford, United Kingdom;
Marsden Fund, New Zealand;
Australian Research Council;
Japan Society for Promotion of Science (JSPS);
the Swiss National Science Foundation (SNSF), Switzerland;
National Research Foundation of Korea (NRF);
Danish National Research Foundation, Denmark (DNRF)

\end{acknowledgments}
\bibliography{mybib}

\end{document}